\documentclass[12pt]{article}
\newwrite\ffile\global\newcount\figno \global\figno=1

\def\writedef#1{}

\input epsf
\def\figin{\epsfcheck\figin}\def\figins{\epsfcheck\figins}
\def\epsfcheck{\ifx\epsfbox\UnDeFiNeD
\message{(NO epsf.tex, FIGURES WILL BE IGNORED)}
\gdef\figin##1{\vskip2in}\gdef\figins##1{\hskip.5in}
\else\message{(FIGURES WILL BE INCLUDED)}%
\gdef\figin##1{##1}\gdef\figins##1{##1}\fi}

\def\figinsert{}
\def\ifig#1#2#3{\xdef#1{fig.~\the\figno}
\writedef{#1\leftbracket fig.\noexpand~\the\figno}%
\figinsert\figin{\centerline{#3}}\medskip\centerline{\vbox{\baselineskip12pt
\advance\hsize by -1truein\center\footnotesize{  Fig.~\the\figno.} #2}}
\bigskip\endinsert\global\advance\figno by1}
\def\endinsert{}

\begin{document}
\baselineskip 18pt
\newcommand{\Tr}{\mbox{Tr\,}}
\newcommand{\beq}{\begin{equation}}
\newcommand{\eeq}{\end{equation}}
\newcommand{\bea}{\begin{eqnarray}}
\newcommand{\eea}[1]{\label{#1}\end{eqnarray}}
\renewcommand{\Re}{\mbox{Re}\,}
\renewcommand{\Im}{\mbox{Im}\,}
\begin{titlepage}

\begin{picture}(0,0)(0,0)
\put(350,0){SHEP-02-26}
\end{picture}

\begin{center}
\hfill
\vskip .4in
{\large\bf
A Stable Supergravity Dual of Non-supersymmetric Glue
}
\end{center}
\vskip .4in
\begin{center}
{\large James Babington, David E. Crooks and Nick Evans}
\footnotetext{e-mail: dc@hep.phys.soton.ac.uk, jrb4@hep.phys.soton.ac.uk,
evans@phys.soton.ac.uk }
\vskip .1in
{\em Department of Physics, Southampton University, Southampton,
S017 1BJ, UK}

\end{center}
\vskip .4in
\begin{center} {\bf ABSTRACT} \end{center}
\begin{quotation}
\noindent We study non-supersymmetric fermion mass and condensate
deformations of the AdS/CFT Correspondence. The 5 dimensional
supergravity flows are lifted to a complete and remarkably simple
10 dimensional background. A brane probe analysis shows that when
all the fermions have an equal mass a positive mass is generated
for all six scalar fields leaving non-supersymmetric Yang Mills
theory in the deep infra-red. We numerically determine the
potential, produced by the background, in the Schroedinger
equation relevant to the study of $O^{++}$ glueballs. The
potential is a bounded well, providing evidence of stability and
for a discrete, confined spectrum. The geometry can also describe
the supergravity background around an (unstable) fuzzy 5-brane.

\end{quotation}
\vfill
\end{titlepage}
\eject
\noindent

\section{Introduction}

The possibility that there is a string description of large $N$ Yang Mills
theory has been speculated on for many years \cite{th}.
The AdS/CFT Correspondence \cite{mald,kleb,ed}
represented the first concrete example of such a duality describing
the ${\cal N}=4$
super Yang Mills (SYM) theory. In this paper we deduce a IIB supergravity
background that describes ${\cal N}=4$ SYM deformed by mass terms for all
the adjoint matter fields leaving pure non-supersymmetric Yang Mills in the
deep infra-red.

In the AdS/CFT Correspondence supergravity fields behave as sources in
the dual gauge theory. Expectation values of field theory operators are
obtained from functional derivatives on the supergravity partition function
with respect to the boundary values of the supergravity fields.
It is therefore a crucial aspect of the correspondence
that the supergravity partition function must be calculable in the presence
of all infinitessimal sources in order that derivatives with respect to
those sources are well defined. In fact, for sources which break the
${\cal N}=4$ theory's conformal symmetry, infinitessimal has no meaning
since they become the only mass scale in the problem. It should therefore
be possible to find gravity duals of deformed versions of the
${\cal N}=4$ SYM theory, including non-supersymmetric theories.

The technology \cite{gppz1, gppz2, freed1, gub}
required to deform the  AdS/CFT Correspondence has
already been developed. In 5d supergravity one must identify the field with
the appropriate symmetry properties to play the role of the field theory
source and determine its action. Solutions of the classical equations of
motion can then be found and are described in the literature as renormalization
group flows. The 5d supergravity backgrounds are often hard to interpret in
terms of the dual field theory. Pilch and Warner \cite{pilch}
have developed a miraculous
ansatz for lifting the 5d solutions to provide the 10d metric and dilaton.
It is then possible, with some work, to solve the field equations for the
remaining supergravity potentials. The resulting solutions are
open to the use of brane probes
\cite{mald,ls,bpp,ejp,more,martelli,abc,prb}
which, using the Dirac Born Infeld action,
provide a direct translation between the gravity background and the
field theory description.

Much of the early work on deformations has concentrated on
supersymmetric theories such as the ${\cal N}=4$ theory on
moduli space \cite{freed2}, ${\cal N}=1$ Leigh Strassler
theory \cite{pw2,ls}, the ${\cal N}=2^*$ \cite{pw,bs2,ep,bpp,ejp}
and ${\cal N}=1^*$ theories \cite{gppz3,ps,pw2}.
More recently interest has turned to non-supersymmetric deformations
of gauge/gravity duals \cite{dz,fred,nsgub,of,epz,nsgub2,bce1}.
Most of these papers have focused  attention  \cite{nsgub,of,epz,nsgub2}
on deformations of more involved ${\cal N}=1$
supersymmetric constructions such as the Maldecena Nunez \cite{mn} and the
Klebanov Strassler \cite{ks} backgrounds. These theories have discrete
vacua and hence supersymmetry breaking perturbations will not result in
an unstable background. The resulting backgrounds are though neccessarily more
complicated than deformations of the ${\cal N}=4$ theory. In \cite{bce1}
we constructed the first complete 10d background of a non-supersymmetric
deformation of ${\cal N}=4$ involving a mass term for a scalar operator.
The resulting field theory and supergravity background shared an instability
in the scalar potential. This highlights one of the most challenging
problems in constructing non-supersymmetric solutions, the need to find
a stable deformation.

In this paper we will deform the AdS/CFT Correspondence by
including a supergravity scalar that is a source for an equal mass
term for each of the four adjoint fermions of ${\cal N}=4$ SYM. We
solve for the 5d supergravity flows numerically. The second order
equations of motion have solutions describing both a mass and a
condensate for the fermion operator. It is easy to see numerically
that if a condensate is present the flows are singular at finite
radius of the AdS space. The mass only solution though is a unique
flow and therefore much harder to numerically study since the
boundary conditions must be arbitrarily fine tuned. Our analysis
though  suggests that this flow may also be  singular in the deep
infra-red. The interpretation of such singularities remains open.
For example the backgrounds describing ${\cal N}=4$ SYM on moduli
space \cite{freed2} are singular but those singularities are
understood to correspond to the presence of D3 branes in the
solution. In the ${\cal N}=2^*$ theory \cite{pw,bs2,ep,bpp,ejp}
the singularities correspond to the divergence of the running gauge 
coupling. On the other hand the backgrounds of Klebanov Strassler
\cite{ks} and Maldecena Nunez \cite{mn} are championed for their
smooth behaviour which is indeed nice. We will not resolve this
issue here, although clearly only after lifting the solution, as
we do, to 10d can one hope to address the physical meaning or
otherwise of singularities. We will find, encouragingly, though
that the singularity does not prevent the background describing a
non-singular glueball spectrum.

The main effort of this paper is to use Pilch and Warner's ansatz
\cite{pilch,pw} and the field equations to construct the full 10d
background appropriate to these 5d solutions. The resulting
background is remarkably simple. The stability of the solution is
then tested using a brane probe. It shows that in the field theory
the 6 scalars have positive masses radiatively induced by the
fermion mass and hence there is no instability to the formation of
a scalar vev (this means that at the 5d supergravity level there
is no instability to the scalar in the 20 of SU(4)$_R$ switching
on). We expect that the SO(4) symmetry acting on the fermions
prevents any other elements of the 5d scalar in the 10 of
SU(4)$_R$ (corresponding to the operators $\lambda_i \lambda_j$)
switching on. At the level of this analysis, the background
appears stable. At first sight it may seem surprising that the 10d
lift has a scalar mass operator present which was not explicitly
introduced at the 5d level. However, this operator is not
represented by a scalar in the 5d supergravity so its presence or
otherwise is not clear at the 5d level. Many of the supersymmetric
deformations \cite{ls,pw,bs2,bpp,ejp,gppz3,pw2} implicitly assume
the presence of this operator in 5d with confirmation only coming
from a brane probe in 10d as we find here. The field theory the
background describes is  ${\cal N}=4$ SYM with masses for all the
matter fields leaving pure non-supersymmetric Yang Mills in the
infra-red. We call this theory Yang Mills$^*$ following the
nomenclature used for supersymmetric deformations of ${\cal N}=4$
SYM.

This theory is hopefully of real use as an approximation to
non-supersymmetric, pure Yang Mills theory. Of course since the
${\cal N}=4$ theory is strongly coupled at all scales, the
deformed theory is strongly coupled at the scale of the mass of
the adjoint matter fields and in this respect differs. This
situation is analogous to the thermalized 5d background of Witten
\cite{ed} which also describes 4d non-supersymmetric Yang Mills in
the infra-red. That theory has been used though to compute
glueball masses \cite{gb} with some success, supporting the use of
these geometries. It will be interesting in the future to compare
the predictions of these two variants to begin to determine the
size of systematic errors induced by the massive matter in each.
As a step in that direction here we numerically compute the
potential in the relevant Schroedinger equation \cite{gb} for the
$O^{++}$ glueballs. For the mass only flow the potential is seen
to be a well, from which we deduce that there is a discrete
spectrum indicating that the geometry indeed describes a confining
gauge theory with a mass gap.

We also note that relative to the thermalized geometry, the Yang
Mills$^*$ deformation is a more systematic approach to obtaining
non-supersymmetric Yang Mills and is more open to the introduction
of quarks (the analysis \cite{karch} of probe D7 branes in anti
de-Sitter (AdS) space appears a particularly fruitful approach).
The thermalization trick would induce masses for the matter fields
too.

The supergravity field we study is also capable of describing an equal
bilinear condensate for each of the four adjoint fermions. As part of
the analysis of the ${\cal N}=1^*$ theory by Polchinski and Strassler \cite{ps}
they
showed that placing a fuzzy D5 brane in AdS induces, asymptotically, precisely
this operator (see \cite{hdcs} for a review of that argument).
These backgrounds then plausibly describe the supergravity
theory induced around a fuzzy 5-brane. A fuzzy 5-brane is not
stable unless there is some force to oppose the potential energy cost of
non-commutative expansion. In the ${\cal N}=1^*$ theory the 5-brane is
polarized by a background 2-form potential (dual to the supersymmetry breaking
mass). An alternative spin was put on the idea in \cite{hdcs} where the
fuzzy expansion was supported by centrifugal force (corresponding to
the presence of a chemical potential in the field theory). In our case there
is no supporting force present and hence it is not surprising that the
brane probe scalar potential is unbounded. The construction is unstable to
the emission of commutative D3 branes.

In the next section we describe the Yang Mills$^*$ deformation in
5d supergravity. In section 3 we describe the oxidation process to
10d and then in section 4 we brane probe the solution. Section 5
discusses the glueball Schroedinger equation for the geometry. The
full background is gathered together in the appendix for ease of
reference.

\section{Deformations in 5d Supergravity}

According to the standard AdS/CFT Correspondence map
\cite{kleb,ed} each supergravity field plays the role of a source
in the dual field theory. The simplest possibility is to consider
non-trivial dynamics for a scalar field in the 5d supergravity
theory. We only allow the scalar to vary in the radial direction
in AdS with the usual interpretation that this corresponds to
renormalization group running of the source. As is standard in the
literature \cite{gppz1,dz} we look for solutions where the metric
is described by

\beq
ds^2 = e^{2 A(r)}dx^\mu dx_\mu + dr^2
\eeq
where $\mu=0..3$ and $r$ is the radial direction in AdS$_5$. The scalar
field has a Lagrangian

\beq
{\cal L} = {1 \over 2} (\partial \lambda)^2 - V(\lambda)
\eeq

There are two independent, non-zero, elements of the Einstein tensor ($G_{00}$
and $G_{rr}$) giving two equations of motion plus there is the
usual equation of motion for the scalar field \cite{gppz1}

\beq \label{e1}
\lambda^{''} + 4 A^{'} \lambda^{'} = {\partial V \over \partial \lambda}
\eeq

\beq  \label{e2}
6 A^{'2} = \lambda^{'2} - 2 V
\eeq

\beq \label{e3}
-3 A^{''} - 6 A^{'2} = \lambda^{'2} + 2 V
\eeq

In fact only two of these equations are independent but it will be useful
to keep track of all of them.

In the large $r$ limit, where the solution will return to AdS$_5$ at first
order and $\lambda \rightarrow 0$ and $V \rightarrow {m^2 \over 2} \lambda^2$,
only the first equation survives with solution

\beq
\lambda = a e^{-\Delta r} + b e^{-(4 - \Delta) r}
\eeq
$a$ and $b$ are constants and

\beq
m^2 = \Delta(\Delta-4)
\eeq
$a$ is interpreted as a source for an operator and $b$
as the vev of that operator since $e^r$ has conformal dimension 1.

If the solution retains some supersymmetry then the potential can
be written in terms of a superpotential

\beq
V = { 1 \over 8} \left| { \partial W \over \partial \lambda} \right|^2 -
{1 \over 3} |W|^2
\eeq
and the second order equations reduce to first order

\beq \label{susyeom}
\lambda^{'} = { 1 \over 2} {\partial W \over \partial \lambda}, \hspace{1cm}
A^{'} = - {1 \over 3} W
\eeq

The deformation we will consider will break supersymmetry completely
and therefore not have such a description.

\subsection{A Fermionic  Operator}

Let us now make a particular choice for the scalar field we will consider.
We take a scalar from the multiplet in the 10 of SO(6).
These operators have been identified \cite{gppz3} as playing the role
of source and vev for the fermionic operator $\psi_i \psi_j$ in the field
theory. In particular we will chose the scalar corresponding to the
operator

\beq \label{op}
{\cal O} = \sum_{i=1}^4 \psi_i \psi_i
\eeq

The potential for the scalar can be obtained from the $N=1^*$ solution
of \cite{gppz3} by setting their two scalars equal (to be precise
one must set their $m = \sqrt{3/ 4} \lambda$ and $\sigma = \sqrt{1 / 4}
\lambda$ to maintain a canonically normalized kinetic term)

\beq \label{pot}
V = - {3 \over 2} \left( 1 + \cosh^2 \lambda \right)
\eeq

In this case $m^2=-3$ and the ultra-violet solutions are

\beq
\lambda = {\cal M} e^{-r} + {\cal K} e^{-3r}
\eeq

The field theory
operators have dimension 1 and 3. Thus in what
follows ${\cal M}=0$ corresponds to a solution with just bi-fermion
vevs while ${\cal K}=0$ corresponds to the purely massive case.

\subsection{Numerical Solutions}

We have not been able to solve the second order equations of
motion analytically but they are easily solved numerically. The
evolution of $\lambda$ as a function of $r$ for a variety of
different initial conditions on $\lambda'$ is shown in Fig 1. The
mass only and condensate only cases are highlighted. The functions
$A(r)$ evolve as $A(r) \sim r$ except in the very deep infra-red.
Note that $\lambda$ typical diverges at finite $r$ with the ${\cal
K}=0$ solution lying on the boundary between solutions that blow
up positively and negatively. In the right hand plot in Fig. 1 we
show solutions close to the mass only solution displaying this
behaviour in the infrared ($r <0$). The uniqueness of the mass
only solution makes it very hard to study in the deep infra-red
because one must arbitrarily fine tune the initial conditions on
$\lambda$. Hence we can not specify the final fate of the mass
only solution in the very deep infra-red although Fig. 1 may
suggest that the solution is diverging.

Whilst analyzing this flow we also analyzed the flow where an
operator corresponding to a vev or a mass for just three of the
four fermions is switched on (in this case the supergravity scalar
potential is given by $V=-3/8 ( 3 + \cosh^2 ( 2 \lambda /
\sqrt{3}) + 4 \cosh (2 \lambda / \sqrt{3})$) . The mass only
solution of this case is one of the ${\cal N} = 1^*$ solutions
\cite{gppz3} which is known analytically. The behaviour of the
second order solutions is very similar to Fig. 1 with the mass
only flow lying, in a similar fashion, between clearly divergent
flows. In that case the analytic, mass only flow does diverge
suggesting further that our mass only flow diverges. As discussed
in the introduction the interpretation of the singularity is a
delicate issue and remains to be determined.
\bigskip

\begin{center}
\hskip-10pt{\lower15pt\hbox{ \epsfysize=2. truein
\epsfbox{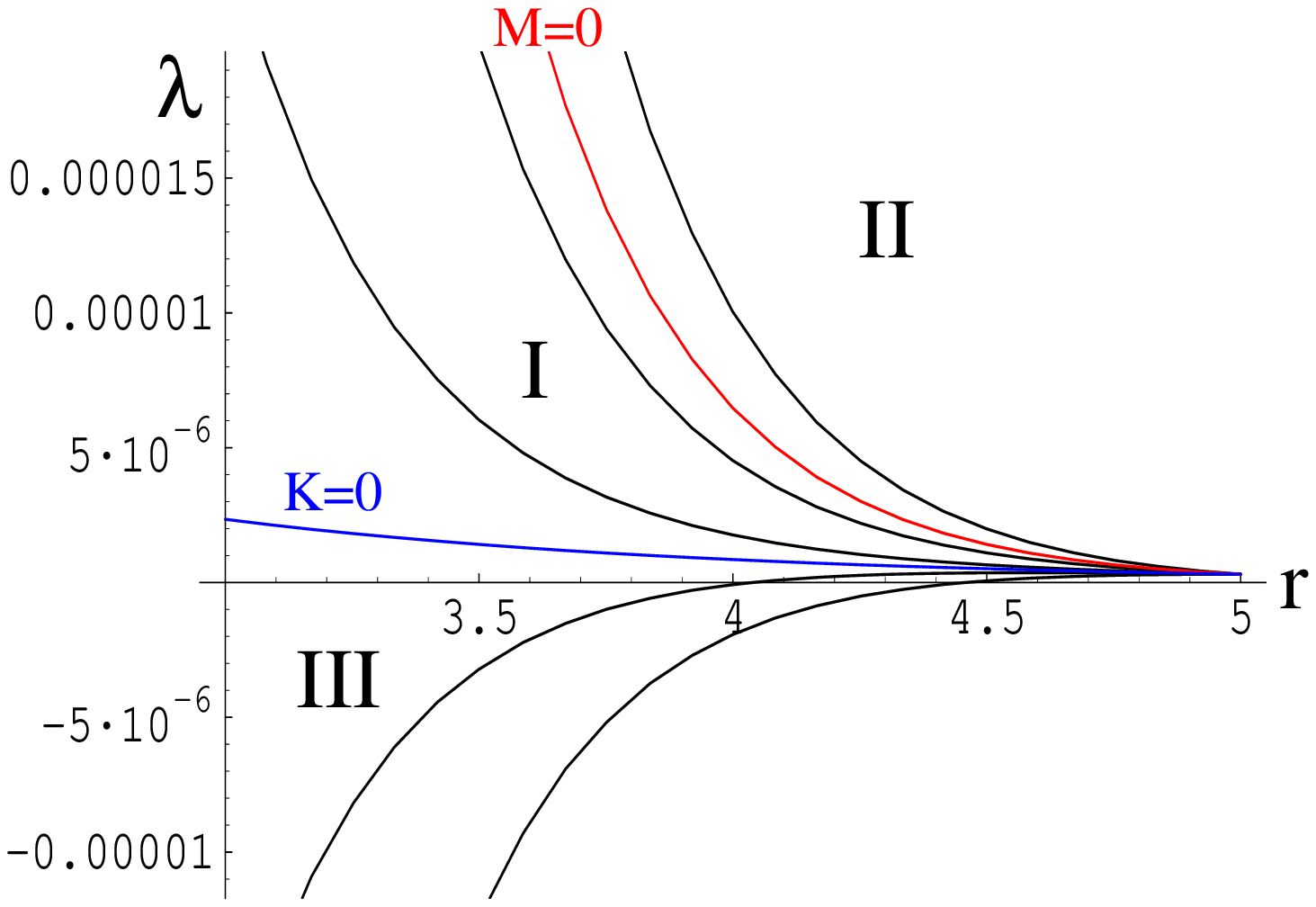}}} \hspace{0.5cm} \hskip-10pt{\lower15pt\hbox{
\epsfysize=2. truein \epsfbox{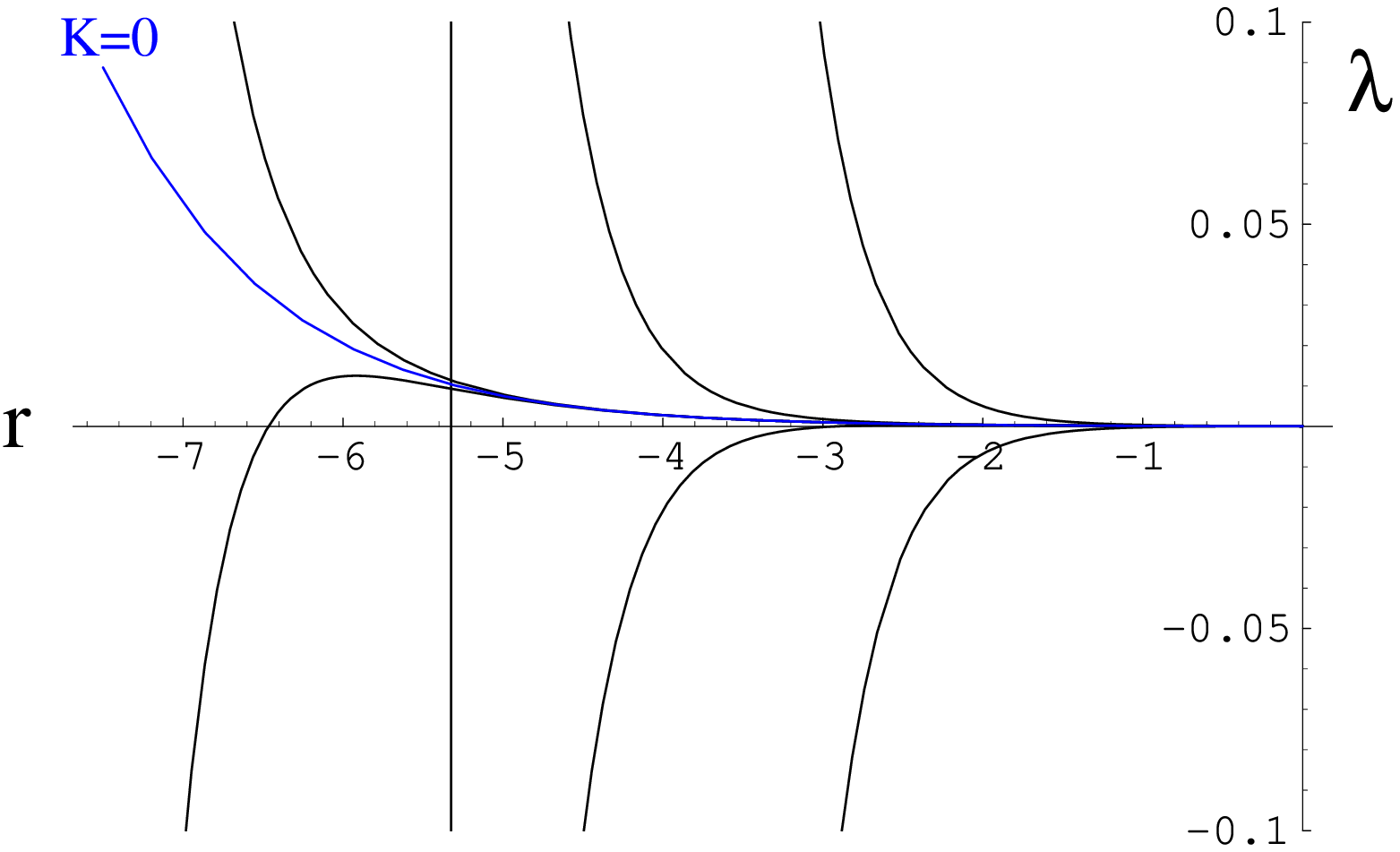}}}
\bigskip

Figure 1: Plots of $\lambda$ vs $r$ for a variety of initial
conditions on $\lambda^{'}$. In the left hand figure the marked
regions correspond to initial conditions: I positive mass and
condensate; II negative mass, positive condensate; III positive
mass, negative condensate. The right hand figure shows a close up
of initial conditions close to the mass only solution in the IR
($r < 0$)
\end{center}

\section{The 10d Background}

To lift the solution to a 10d background requires us to find a solution of the
full set of IIB supergravity equations of motion. As was found in \cite{pw2,pw}
where a fermion mass term was introduced in a supersymmetric context
all the supergravity fields will be non-zero. We first summarize the
field equations taken from \cite{shw} \medskip

\noindent
$\bullet$ The Einstein equations:\medskip

\beq
R_{MN}=T^{(1)}_{MN}+T^{(3)}_{MN}+T^{(5)}_{MN}
\eeq
where the energy momentum tensor contributions from the dilaton,
2-form potential and 4-form potential are given by

\beq
T^{(1)}_{MN}= P_MP_N{}^*+P_NP_M{}^*
\eeq

\beq
T^{(3)}_{MN}=
       {1\over 8}(G^{PQ}{}_MG^*_{PQN}+G^{*PQ}{}_MG_{PQN} -
        {1\over 6}g_{MN} G^{PQR}G^*_{PQR})
\eeq

\beq
T^{(5)}_{MN}= {1\over 6} F^{PQRS}{}_MF_{PQRSN}
\eeq\medskip

The dilaton is written in unitary gauge where

\beq
P_M= f^2\partial_M B, \hspace{1cm}  Q_M= f^2   {\rm Im} (
B\partial_MB^*),
 \hspace{1cm}
f= {1\over (1-BB^*)^{1/2}} \eeq

The more familiar dilaton-axion field is given by

\beq \label{ax}
a + i e^\Phi = i {(1-B) \over (1+B)}
\eeq
and the 3-form field strength is defined by

\beq
G_{(3)}= f(F_{(3)}-BF_{(3)}^*)
\eeq\medskip

\noindent
$\bullet$ The Maxwell equations:\medskip

\beq
(\nabla^P-i Q^P) G_{MNP}= P^P G^*_{MNP}-{2\over 3}\,i\,F_{MNPQR}
G^{PQR}
\eeq

\beq
(\nabla^M -2 i Q^M) P_M= -{1\over 24} G^{PQR}G_{PQR}
\eeq\medskip

\noindent
$\bullet$ The self-dual equation:\medskip

\beq
F_{(5)}= \star F_{(5)}
\eeq \medskip

\noindent
$\bullet$ Bianchi identities:

\beq
F_{(3)}= dA_{(2)}, \hspace{0.5cm}
d F_{(5)}= -{1\over 8}\,{\rm Im}( F_{(3)}\wedge
F_{(3)}^*)
\eeq

\subsection{The UV limit}

Let us first concentrate on lifting the ultra-violet
($r \rightarrow \infty$) limit
of the 5d flow. The supergravity scalar lifts to the 2-form potential in 10d
\cite{neu}.
To determine its form we can use the group theory technique
in \cite{ps,hdcs}.

Parameterize the 6d space perpendicular to the D3 branes of the
construction as

\beq
z_1 = { w^1 + i y^1 \over \sqrt{2}}, \hspace{0.5cm} z^2 = { w^2 + i
y^2 \over \sqrt{2}}, \hspace{0.5cm} z_3 = { w^3 + i y^3 \over
\sqrt{2}}
\eeq

Under the SO(2) rotation subgroups of the 6 dimensional representation
of SU(4),
$z^i \rightarrow e^{i \phi_i} z_i$, the 4 dimensional representation
transforms as

\beq
\begin{array}{c}
\lambda_1 \rightarrow e^{i ( \phi_1 - \phi_2 - \phi_3)/2} \lambda_1,
\hspace{0.5cm}
\lambda_2 \rightarrow e^{i ( -\phi_1 + \phi_2 - \phi_3)/2} \lambda_2, \\
\\
\lambda_3 \rightarrow e^{i ( -\phi_1 - \phi_2 + \phi_3)/2} \lambda_3,
\hspace{0.5cm}
\lambda_4 \rightarrow e^{i ( \phi_1 + \phi_2 + \phi_3)/2} \lambda_4.
\end{array}
\eeq

We can thus construct a 3-form field strength with the symmetry properties
of a fermion mass or condensate

\beq
\langle \lambda_1 \lambda_1 \rangle dz^1 \wedge d\bar{z}^2 \wedge d\bar{z}^3
+ \langle \lambda_2 \lambda_2 \rangle d\bar{z}^1 \wedge dz^2 \wedge d\bar{z}^3
+ \langle \lambda_3 \lambda_3 \rangle d\bar{z}^1 \wedge d\bar{z}^2 \wedge dz^3
+ \langle \lambda_4 \lambda_4 \rangle dz^1 \wedge dz^2 \wedge dz^3
\eeq
setting all the operators equal gives

\beq
F_{(3)} =   dw^1 \wedge dw^2 \wedge dw^3
+ i  dy^1 \wedge dy^2 \wedge dy^3
\eeq

It will be useful to write the 6d space in terms of two
$S^2$ (with metrics $d\Omega_{\pm}^2 = d \theta_\pm^2 +
\cos^2 \theta_\pm d \phi_\pm^2$) corresponding to the $w$ and $y$ spaces,
a radial direction $r$,
and an angular coordinate between the spheres, $\alpha$. The
appropriate 2-form is then

\beq
A_{(2)} =   \cos^3 \alpha \cos \theta_+
d \theta_+ d \wedge \phi_+
+ i \sin^3 \alpha  \cos \theta_-  d \theta_- \wedge d \phi_-
\eeq

A survey of the field equations reveals  that only
the 2-form's Maxwell equation is of leading order in the perturbing field
$\lambda$. We find that

\beq
A_{(2)} = 2 \lambda (i \cos^3 \alpha \cos \theta_+
d \theta_+ \wedge d \phi_+
- \sin^3 \alpha  \cos \theta_-  d \theta_- \wedge d \phi_-)
\eeq
indeed reproduces the asymptotic form of the 5d field equation

\beq
\lambda^{''} + 4 \lambda^{'} = - 3
\eeq
when substituted into that Maxwell equation.

The ultra-violet solution for $A_{(2)}$ provides a
useful check at each stage of the computation of
the full lift which we come to next.

\subsection{The Metric}

Pilch and Warner \cite{pilch,pw2} have provided an ansatz for the
lift of a 5d supergravity flow to 10d (note that, although they
study supersymmetric flows, their ansatz is not restricted to the
supersymmetric solution of the second order equations of motion).
In particular the ansatz provides us with the metric and dilaton.
We can find the lifts we want as a limit of the metrics in
\cite{pw2}; that lift is of the ${\cal N}=1^*$ GPPZ flows
\cite{gppz3} in which three of the fermions are given a mass and
the fourth develops a bilinear condensate. Setting these scalars
equal  (again to be precise one must set their $m = \sqrt{3/ 4}
\lambda$ and $\sigma = \sqrt{1 / 4} \lambda$ to maintain a
canonically normalized kinetic term) gives the metric we require

\beq
ds_{10}^2 = \xi^{1 \over 2} ds_{1,4}^2+ \xi^{-{3 \over 2}}  ds_{5}^2
\eeq

The metric of the deformed five sphere in their coordinates
($u^i, v^i, i=1..3$) is given by
\beq
ds_5^2 =c^2 du^{i}du_{i}-4 s^2 u.v du^{i}dv_{i}+c^2 dv^{i}dv_{i}
+c^2s^2 d(u.v)^2
\eeq
where

\beq
c= \cosh \lambda, \hspace{1cm} s = \sinh \lambda
\eeq
This metric is subject to the constraint

\beq
u^2 + v^2 =1
\eeq

\noindent The warp factor is given by

\beq
\xi^2=c^4 -4 s^4 (u.v)^2
\eeq

We must move to more appropriate coordinates for our problem. The
metric can be diagonalized by the change of coordinates

\beq
U_{\pm}^{i}={1 \over \sqrt{2}} [u^{i} \pm v^{i}]
\eeq

\beq
ds_5^2 =(c^2 -s^2 [U_{+}^{2} -U_{-}^{2}]) dU_{+}^{i}dU_{+}^{i}
+(c^2 +s^2 [U_{+}^{2} -U_{-}^{2}]) dU_{-}^{i}dU_{-}^{i}
+4c^2s^2 U_{+}^{i}U_{+}^{j}dU_{+}^{i}dU_{+}^{j}
\eeq

The constraint can now be applied using the coordinates used in
the UV limit above ($r,\alpha$ and two $S^2$ parameterized by
$\theta_\pm, \phi_\pm$)

\beq \begin{array}{c}
U_{+}^{1}= \cos\theta_{+} \cos \phi_{+} \cos \alpha, \hspace{0.5cm} U_{+}^{2}
= \sin\theta_{+} \cos \phi_{+} \cos \alpha, \hspace{0.5cm}U_{+}^{3}
= \sin\theta_{+}  \cos \alpha
\\
 \\
U_{-}^{1}= \cos\theta_{-} \cos \phi_{-} \sin \alpha, \hspace{0.5cm}
U_{-}^{2}= \sin\theta_{-} \cos \phi_{-} \sin \alpha, \hspace{0.5cm}
U_{-}^{3}= \sin\theta_{-}  \sin \alpha
\end{array}
\eeq

\noindent The metric then takes the form

\beq ds_5^2 =\cos^2 \alpha (c^2 -s^2 \cos 2\alpha) d\Omega_{+}^2 +
\sin^2 \alpha(c^2 +s^2 \cos 2\alpha) d\Omega_{-}^2 +\xi^2
d\alpha^2 \eeq

\noindent and the warp factor becomes

\beq
\xi^2=c^4 -s^4 \cos^2 2\alpha
\eeq

\subsection{The Ricci Tensor}

The calculation of the Ricci tensor is carried out by computer and
the second order 5d flow equations for the scalar field are used throughout
to simplify the expressions. The results are lengthy but we note that
the non-zero components of the Ricci tensor are

\beq
R_{00}=R_{11}=R_{22}=R_{33},\hspace{0.7cm}
R_{rr},\hspace{0.7cm} R_{\alpha\alpha}, \hspace{0.7cm}
R_{r \alpha}=R_{\alpha r}
, \hspace{0.7cm}
R_{66}=R_{77}, \hspace{0.7cm} R_{88}=R_{99}
\eeq

\subsection{The Dilaton}

The dilaton can again be extracted from \cite{pw2} where they provide

\beq
{\cal M} ~=~ {\cal S}{\cal S}^T~ = ~
{ 1 \over \xi} \left( \begin{array}{cc} \cosh^2 \lambda &
  \sinh^2 \lambda \cos 2 \alpha\\
 \sinh^2 \lambda \cos 2 \alpha& \cosh^2 \lambda
\end{array}
\right)
\eeq
where (in unitary gauge)

\beq
{\cal S} = f \left( \begin{array}{cc} 1 & B \\
B^* & 1
\end{array}
\right), \hspace{1cm} f = { 1 \over ( 1 - |B|^2)^{1/2}}
\eeq
we thus find

\beq
f = { 1 \over \xi^{1/2}} \sqrt{\cosh^2 \lambda + \xi \over 2}, \hspace{1cm}
B = {  \sinh^2 \lambda \cos 2 \alpha \over  \cosh^2 \lambda + \xi}
\eeq

Note that $B$ is a real function and therefore
from (\ref{ax}) the axion is zero for this flow. The $r$ dependence
implies the gauge coupling runs although finding the correct coordinate
system to match it to the gauge theory will be difficult.

\subsection{The 2-form and 4-form Potentials}

We now move on to determining the potentials in the solution.
Motivated by the UV limit we make an ansatz for the 2-form potential
of the form

\beq
A_{(2)}=i A_{+} (\lambda(r),\alpha) \cos^3 \alpha \, \cos \theta_{+}
d \theta_{+} \wedge d \phi_{+} - A_{-}(\lambda(r),\alpha)
\sin^3 \alpha \, \cos \theta_{-} d \theta_{-} \wedge d \phi_{-}
\eeq

\noindent where $A_{+}$ and $A_{-}$
are arbitrary functions
that become $\lambda(r)$ in the UV.

The non-vanishing components of the three form energy momentum tensor
are then

\beq
T^{\, \,0 }_{(3) 0}=T^{\, \,1}_{(3) 1}=T^{\, \,2 }_{(3) 2}=T^{\, \,3 }_{(3) 3}
=-\frac{1}{8} [\mathcal{A}+\mathcal{B}+\mathcal{C}+\mathcal{D}]
\eeq

\beq
T^{\, \,r}_{(3) r}=\frac{1}{2}[\mathcal{A}+\mathcal{C}] +T^{\, \,0}_{(3) 0}
\eeq

\beq
T^{\, \,\alpha}_{(3)\alpha}=\frac{1}{2}[\mathcal{B}+\mathcal{D}] +T^{\, \,0}_{(3) 0}
\eeq

\beq
T^{\, \,6 }_{(3) 6}=T^{\, \,7}_{(3) 7}=\frac{1}{2}[\mathcal{A}+\mathcal{B}] +T^{\, \,0}_{(3)0}
\eeq

\beq
T^{\, \, 8 }_{(3) 8}=T^{\, \, 9 }_{(3) 9}=\frac{1}{2}[\mathcal{C}+\mathcal{D}] +T^{\, \, 0 }_{(3) 0}
\eeq

\beq
T^{\, \,r}_{(3)\alpha}=\frac{1}{2}[G^{r67} G_{\alpha 67}+G^{r89}G_{\alpha  89}]
\eeq

\noindent where the functions $\mathcal{A}$, $\mathcal{B}$, $\mathcal{C}$
and $\mathcal{D}$ are given by

\beq \begin{array}{c}
\mathcal{A}=g^{rr}g^{66}g^{77}|G_{r67}|^2 ,\hspace{0.5cm}
\mathcal{B}=g^{\alpha \alpha}g^{66}g^{77}|G_{\alpha 67}|^2,\\
\\
\mathcal{C}=g^{rr}g^{88}g^{99}|G_{r89}|^2,\hspace{0.5cm}
\mathcal{D}=g^{\alpha \alpha}g^{88}g^{99}|G_{\alpha 89}|^2
\end{array}
\eeq \medskip

We now turn to the 4-potential. The self duality condition of the
five form field strength is satisfied by construction using the
ansatz

\beq
F_{(5)}=\mathcal{F} +\star \mathcal{F}, \hspace{1cm} \mathcal{F}
= dx^{0}\wedge dx^{1}\wedge dx^{2}\wedge dx^{3}\wedge d\omega(r,\alpha)
\eeq
The non-vanishing components of the five form energy momentum tensor are

\beq
T^{\, \,0 }_{(5) 0}=T^{\, \,1}_{(5) 1}=T^{\, \,2}_{(5) 2}
=T^{\, \,3 }_{(5) 3}=-T^{\, \,6}_{(5) 6}=-T^{\, \,7 }_{(5) 7}
=-T^{\, \, 8}_{(5) 8}=-T^{\, \, 9}_{(5) 9}=\mathcal{X} +\mathcal{Y}
\eeq

\beq
T^{\, \, r}_{(5) r}=-T^{\, \, \alpha}_{(5)\alpha}=\mathcal{X} -\mathcal{Y}
\eeq

\noindent
where the functions $\mathcal{X}$ and $\mathcal{Y}$ are given by

\beq
\mathcal{X} = \frac{1}{2}g^{00}g^{11}g^{22}g^{33}g^{rr} (\frac{\partial \omega}{\partial r})^2, \hspace{1cm}
\mathcal{Y} = \frac{1}{2}g^{00}g^{11}g^{22}g^{33}g^{\alpha \alpha} (\frac{\partial \omega}{\partial \alpha})^2
\eeq

\noindent and

\beq
T^{\, \,r}_{(5) \alpha}=\frac{1}{2}g^{00}g^{11}g^{22}g^{33}g^{rr} (\frac{\partial \omega}{\partial \alpha}\frac{\partial \omega}{\partial r})
\eeq \medskip

To solve the supergravity equations we need to disentangle the
contributions from the 2-form potential and the 4-form.
Furthermore, we need to separate the function $A_{+}$ from
$A_{-}$. We achieve this with the following combinations in which
the 4-form cancels

\beq
R^{7}_{\,\, 7}-R^{9}_{\,\, 9}+2R^{r}_{\,\, r}+2R^{\alpha}_{\,\,\alpha }-
2T^{\, \,r}_{(1) r}-2T^{\, \,\alpha}_{(5) \alpha}=\mathcal{A} + \mathcal{B}
\eeq

\beq
R^{9}_{\,\, 9}-R^{7}_{\,\, 7}+2R^{r}_{\,\, r}+2R^{\alpha}_{\,\,\alpha }-
2T^{\, \,r}_{(1) r}-2T^{\,\,\alpha}_{(5) \alpha}=\mathcal{C} + \mathcal{D}
\eeq
Since $\lambda$ is the only function of $r$ in the solution we can separate
out the pieces proportional to $\lambda^{'2}$ and those not, to
distinguish between, for example, $\mathcal{A}$ and $\mathcal{B}$.
A lengthy  calculation, in which the 5d field equations
are used repeatedly to simplify expressions,  yields the simple result

\beq
A_{\pm}(r,\alpha)=\frac{\sinh 2\,\lambda }
{\cosh^2\lambda \pm \cos 2\alpha \, \sinh^2 \lambda}
\eeq

The remaining function in the 4-form potential
$\omega(r,\alpha)$ can now be found using the equations

\beq
R^{0}_{\,\, 0}+R^{r}_{\,\, r}-T^{\, \,r}_{(1) r}-T^{\, \,r}_{(3) r}-
T^{\, \, 0}_{(3) 0}=\mathcal{X}
\eeq

\beq
R^{0}_{\,\, 0}-R^{r}_{\,\, r}+T^{\, \,r}_{(1) r}+T^{\, \,r}_{(3) r}
-T^{\, \,0}_{(3) 0}=\mathcal{Y}
\eeq\medskip

In fact there is no angular dependence in $\omega$ and we find again the
simple result

\beq
\frac{\partial \omega}{\partial \alpha}=0, \hspace{1cm} \frac{\partial \omega}{\partial r}= e^{4A(r)} V(r)
\eeq

\noindent And hence

\beq
\omega(r)=e^{4A(r)} A'(r)
\eeq

This completes the solution. The remaining equations of motion act as a
check of the solution.

\section{Brane Probing}

The most successful technique for connecting backgrounds and their
dual field theories has been brane probing
\cite{mald,bpp,ejp,ls,more,martelli,abc,prb} which converts the
background to the U(1) theory on the probe's surface. We thus
substitute the background into the Born-Infeld action which, since
the 2-form field is entirely orthogonal to the probe directions,
takes the form \cite{bpp,ejp}

\begin{equation} \label{BI}
S_{probe}=-\tau_3\int_{\mathcal{M}_4}d^4x
\det[G^{(E)}_{ab} + 2 \pi \alpha' e^{- \Phi/2} F_{ab}]^{1/2}
+ \mu_3\int_{\mathcal{M}_4} C_4,
\end{equation}

\noindent where $C_{(4)}$ is the pull back of
the 4-form potential on to the brane
which corresponds here to the function $w$ above.
The resulting scalar potential is given by

\beq
V_{probe} = e^{4A} \left[\xi - A^{'} \right]
\eeq \bigskip

\begin{center}
\hskip-10pt{\lower15pt\hbox{
\epsfysize=2.5 truein \epsfbox{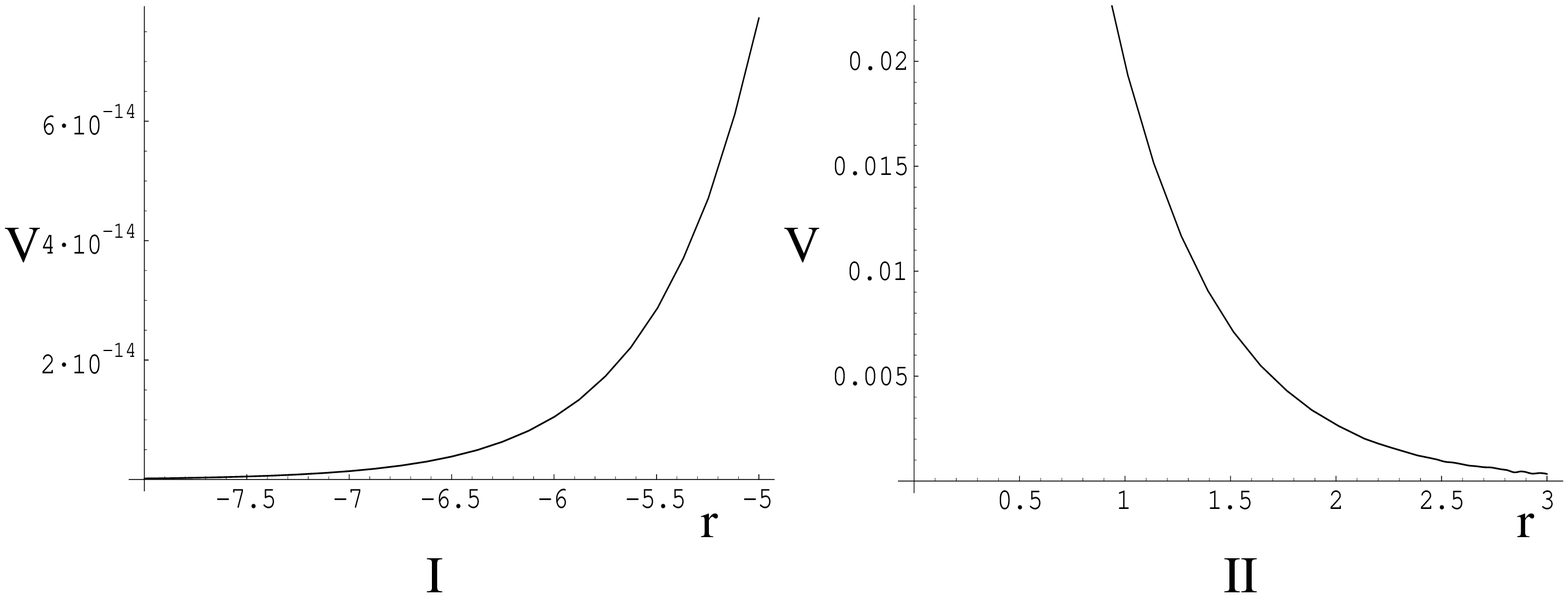}}}\medskip

Figure 2:
Plots of the probe potential in the infra-red for the mass only solution
showing the stability of the solution (I)
and the condensate only solution which is unstable (II).
\end{center}

\subsection{Yang Mills$^*$ Boundary Conditions}

It is illuminating to evaluate this potential at leading order in the
ultra-violet with

\beq
\lambda = {\cal M}  e^{- r}+ ..., \hspace{1cm} A = r + ...
\eeq
We find

\beq
V = {\cal M} ^2 e^{2 r} + ...
\eeq

Remembering that $e^r$ has conformal dimension of mass this is an
equal mass term for each of the 6 scalar fields. The field theory
at large scalar vevs is bounded suggesting the set up is stable.
Note that the potential's dependence on the angle $\alpha$ in
$\xi$ is subleading in the ultra-violet. The infra-red behaviour
can be found numerically by solving (\ref{e1}) and (\ref{e2}) and
a sample plot is shown in Fig 2. The potential is largely
independent of $\alpha$ in the infra-red too. The plot supports
the hypothesis that the scalar potential pins the probe at the
origin of the space.

This background therefore appears to be the dual of a stable non-supersymmetric
gauge theory in which all the adjoint matter fields are massive. The deep
infra-red physics is pure Yang Mills.

\subsection{Fuzzy Sphere Boundary Conditions}

Alternatively if we look at the other possible asymptotic solution

\beq
\lambda = {\cal K}  e^{- 3 r}+ ..., \hspace{1cm} A = r + ...
\eeq
We find

\beq
V = {\cal K}^2 e^{-2 r} + ...
\eeq
a condensate leaves a runaway potential. Again the infra-red
behaviour can be found numerically (Fig 2) and shows the same
behaviour as the asymptotic solution. This configuration, which
asymptotically looks like the $F_{(3)}$ field we would expect a D5 in AdS
to generate, is unstable to the emission of probe like D3 branes. This
is not surprising since there is no force supporting the expansion of the
D3s into a fuzzy D5 brane.

The other possible solutions with both a mass and a condensate present
interpolate between the two forms of solution we've seen. In the infra-red
they are unstable whilst in the ultra-violet
the mass term dominates. In between there
is a minimum of the probe potential. However, given the instability of the
core structure there is probably little physics associated with this minimum.

\section{Glueballs and Confinement}

We can make an initial investigation of the infra-red properties
of the gauge theory described by our geometry as follows. The
$O^{++}$ glueballs of the theory have been identified \cite{gb}
with excitations of the dilaton field of the form

\beq \delta \Phi = \psi(r) e^{-i k x}, \hspace{1cm} k^2 = -M^2
\eeq
This deformation must be a solution of the 5d dilaton field
equation

\beq \partial_\mu( \sqrt{-g} g^{\mu \nu} \partial_\nu) \delta \Phi
= 0 \eeq
If we make the change of coordinates ($r \rightarrow z$) such that

\beq {d z \over dr} = e^{2 A} \eeq
and rescale

\beq \psi \rightarrow e^{-3 A/2} \psi\eeq
Then the dilaton field equation takes a Schroedinger form

\beq ( - \partial_z^2 + V(z)) \psi(z) = M^2 \psi(z)  \eeq
where

\beq V= {3 \over 2} A^{''} + {9 \over 4} (A^{'})^2 \eeq \medskip

In these coordinates the equations of motion become

\beq  \lambda^{''} + 3 \lambda^{'} A^{'} = e^{2 A} {\partial V
\over \partial \lambda} \eeq

\beq 6 A^{'2} =  \lambda^{'2} - 2 e^{2 A} V \eeq

It is now a simple matter to solve these equations numerically
with UV boundary conditions

\beq  \lambda \simeq {\cal M} z + {\cal K} z^3, \hspace{1cm} A
\simeq - \log z \eeq

By suitably fine tuning the boundary conditions close to the mass
only solution we can numerically determine the potential relevant
to the Schroedinger equation. We display the results in Fig. 3.
Note that if there is any fermion condensate present then the
potential is unbounded but as we tune onto the mass only solution
we find a bounded potential well. This behaviour is again
analogous to that in the case where a mass is given to three of
the four fermions as discussed in section 2.2 - there the mass
only case is known analytically and the potential is known to be
fully bounded \cite{ma}.

It is clear from the potential that there are stable, discrete
glueball states in the Yang Mills$^*$ geometry. This immediately
implies that the theory is confining and has a mass gap
which is encouraging. The
boundedness  of the well also provides further support for the
claim that the geometry and field theory are stable. The
instability of solutions with a condensate for the fermion
operator matches nicely with the instability of the brane probe in
these cases. We leave further investigation of the infra-red
dynamics for future work.

\begin{center}
\hskip-10pt{\lower15pt\hbox{ \epsfysize=2. truein
\epsfbox{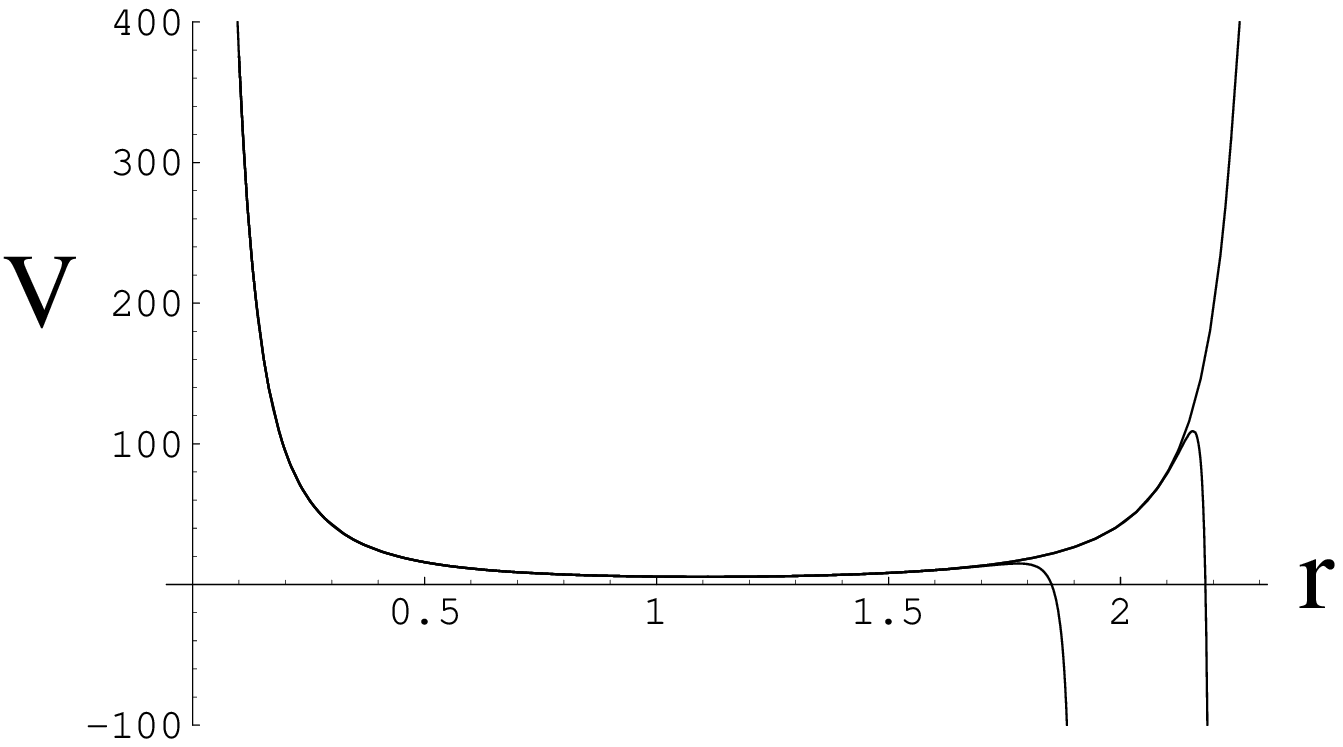}}}\medskip

Figure 3: Plots of the Schroedinger potential relevant to the
$O^{++}$ glueballs for three flows progressively fine tuned
towards the mass only (Yang Mills$^*$) geometry. The presence of a
condensate leads to an unstable potential, as can be seen at the
right hand side of the well, but as it is reduced a clearly
bounded potential well emerges.
\end{center}

\section{Summary}

In this paper we have studied deformations of the AdS/CFT
Correspondence which are bi-fermion masses or condensates in the
field theory. The major challenge has been to lift the solutions
to a complete 10d IIB supergravity background. The resulting
background, summarized in the appendix, is surprisingly simple. We
have brane probed the solution in order to study the field theory
scalar potential. For the mass only solution the probe potential
is stable and the scalars massive. This theory is therefore
non-supersymmetric Yang Mills theory in the infra-red. We hope
that this geometry will provide a new tool for studying Yang Mills
theory. As a first step in this direction we have determined the
Schroedinger potential relevant to the study of $O^{++}$ glueballs
in the geometry, which is a bounded well. One can hence deduce
that the geometry indeed describes a confined spectrum in the
infra-red. It should also be possible in the future to include
probe D7 branes in the geometry and study the fermionic quark
potential for chiral symmetry breaking. Whether this Yang
Mills$^*$ theory is a good approximation to pure Yang Mills
remains to be seen.

Any solution with a fermion condensate present generates an
unstable probe and glueball potential. The asymptotic form of the
solution suggests there is a D5 brane in the core of the geometry.
We have interpreted these solutions as the geometries around a
fuzzy D5 brane with no force supporting the non-commutative
expansion.

\vspace{2cm}

\noindent {\bf Acknowledgements:} We are grateful to Alberto Zaffaroni,
Michela Petrini
and Clifford Johnson for comments on the manuscript, and Nick Warner
for discussions.
NE is grateful to PPARC for the support
of an Advanced Fellowship. JRB and DC are grateful to PPARC for the support
of studentships.
\vspace{1cm}

\newpage

\section{Appendix - Summary of the Yang Mills$^*$ Geometry}

The geometry is

\beq
ds_{10}^2 = \xi^{1 \over 2} ds_{1,4}^2+ \xi^{-{3 \over 2}}  ds_{5}^2
\eeq\vspace{-0.9cm}

\beq
\xi^2=c^4 -s^4 \cos^2 2\alpha, \hspace{1cm} c= \cosh \lambda, \hspace{1cm}
s = \sinh \lambda
\eeq\vspace{-0.9cm}

\beq
ds_{1,4}^2 = e^{2 A} dx_{//}^2 + dr^2
\eeq\vspace{-0.9cm}

\beq ds_5^2 =\cos^2 \alpha (c^2 -s^2 \cos 2\alpha) d\Omega_{+}^2 +
\sin^2 \alpha(c^2 +s^2 \cos 2\alpha) d\Omega_{-}^2 +\xi^2
d\alpha^2 \eeq \beq d \Omega_\pm^2 = d \theta_\pm^2 + \cos^2
\theta_\pm d \phi_\pm^2 \eeq

\noindent The dilaton in unitary gauge is described by the functions

\beq
f = { 1 \over \xi^{1/2}} \sqrt{\cosh^2 \lambda + \xi \over 2}, \hspace{1cm}
B = {  \sinh^2 \lambda \cos 2 \alpha \over  \cosh^2 \lambda + \xi}
\eeq

\noindent The 2-form potential is given by

\beq
A_{(2)}=i A_{+}(r,\alpha) \cos^3 \alpha \, \cos \theta_{+}
d \theta_{+} \wedge d \phi_{+} -  A_{-}(r,\alpha)
\sin^3 \alpha \, \cos \theta_{-} d \theta_{-} \wedge d \phi_{-}
\eeq

\beq
A_{\pm}(r,\alpha)=\frac{\sinh 2\,\lambda }
{\cosh^2\lambda \pm \cos 2\alpha \, \sinh^2 \lambda}
\eeq

\noindent The 4-form potential is given by

\beq
F=\mathcal{F} +\star \mathcal{F}, \hspace{1cm} \mathcal{F}
= dx^{0}\wedge dx^{1}\wedge dx^{2}\wedge dx^{3}\wedge d\omega
\eeq

\beq
\omega(r)=e^{4A(r)} A'(r)
\eeq

\noindent The functions $A$ and $\lambda$ are solutions of

\beq
\lambda^{''} + 4 A^{'} \lambda^{'} = {\partial V \over \partial \lambda}
\eeq

\beq
6 A^{'2} = \lambda^{'2} - 2 V
\eeq

\noindent with

\beq
V = - {3 \over 2} \left( 1 + \cosh^2 \lambda \right)
\eeq


\newpage

\begin{thebibliography}{99}

\bibitem{th} G. 'tHooft, Nucl. Phys. {\bf B72} (1974) 461.

\bibitem{mald} J. Maldacena, Adv. Theor. Math. Phys. {\bf 2} (1998) 231,
hep-th/9711200.
\bibitem{kleb} S.S. Gubser, I.R. Klebanov and A.M. Polyakov,
Phys. Lett. {\bf B428} (1998) 105, hep-th/9802109.
\bibitem{ed} E. Witten, Adv. Theor. Math. Phys. {\bf 2}
(1998) 253, hep-th/9802150.
\bibitem{gppz1} L. Girardello, M. Petrini, M. Porrati and A. Zaffaroni,
JHEP {\bf 9812} (1998) 022, hep-th/9810126.
\bibitem{gppz2} L. Girardello, M. Petrini, M. Porrati and A. Zaffaroni,
JHEP {\bf 9905} (1999) 026, hep-th/9903026.
%
\bibitem{freed1} D. Z. Freedman, S. S. Gubser, K. Pilch and
N. P. Warner, Adv. Theor. Math. Phys. {\bf 3} (1999) 363, hep-th/9904017.
\bibitem{gub} S.S. Gubser, Adv. Theor. Math. Phys. {\bf 4} (2002) 679,
hep-th/0002160.
\bibitem{pilch} A. Khavaev, K. Pilch and N. P. Warner,
Phys. Lett. {\bf B487} (2000) 14, hep-th/9812035.
\bibitem{ls} C.V. Johnson, K.J. Lovis, D.C. Page  JHEP {\bf 0105} (2001) 036,
 hep-th/0011166; JHEP {\bf 0110} (2001) 014,
hep-th/0107261.
%
\bibitem{bpp} A. Buchel, A.W. Peet and J. Polchinksi,
Phys. Rev. {\bf D63} (2001) 044009, hep-th/0008076.
%
\bibitem{ejp} N. Evans, C.V. Johnson and M. Petrini,
JHEP {\bf 0010} (2000) 022,  hep-th/0008081.
\bibitem{more}
J. Babington, N. Evans, J, Hockings, JHEP {\bf 0107} (2001) 034,
 hep-th/0105235.
\bibitem{martelli} J. P. Gauntlett, N. Kim, D. Martelli
and D. Waldram, Phys. Rev. {\bf D64} (2001) 106008, hep-th/0106117.
\bibitem{abc}
F.Bigazzi, A.L. Cotrone, A. Zaffaroni,  Phys. Lett. {\bf B519}
(2001) 269, hep-th/0106160.
%
\bibitem{prb} R. de Mello Koch, A. Paulin-Campbell, J.P. Rodrigues
Phys. Rev. {\bf D60} (1999) 106008, hep-th/9903029; Nucl.
Phys. {\bf B559} (1999) 143, hep-th/9903207.
\bibitem{freed2} D. Z. Freedman, S. S. Gubser, K. Pilch and
N. P. Warner, JHEP {\bf 0007} (2000) 038, hep-th/9906194.
\bibitem{pw2}  K. Pilch, N.P. Warner, Adv. Theor. Math. Phys. {\bf 4} (2002)
627, hep-th/0006066.
\bibitem{pw} K. Pilch and N. P. Warner, Nucl.Phys. {\bf B594} (2001) 209,
hep-th/0004063.
\bibitem{bs2} A. Brandhuber and K. Sfetsos, Phys. Lett. {\bf B488} (2000) 373,
 hep-th/0004148.
\bibitem{ep} N. Evans and M. Petrini,
Nucl. Phys. {\bf B592} (2001) 129,  hep-th/0006048.
%
\bibitem{gppz3} L. Girardello, M. Petrini, M. Porrati and A. Zaffaroni,
Nucl.Phys. B569 (2000) 451, hep-th/9909047.
%
\bibitem{ps} J. Polchinski, M.J. Strassler, {\it The String Dual of
a Confining Four-dimensional Gauge Theory}, hep-th/0003136.
%
\bibitem{dz} J. Distler and F. Zamora, Adv. Theor. Math. Phys. {\bf 2} (1998)
1405, hep-th/9810206.
%
\bibitem{fred}F. Zamora,  JHEP {\bf 0012} (2000) 021, hep-th/0007082.
%
\bibitem{nsgub} S.S. Gubser, A.A. Tseytlin, M.S. Volkov, JHEP {\bf 0109}
(2001) 017, hep-th/0108205.
\bibitem{of} O. Aharony, E. Schreiber, J. Sonnenschein, JHEP {\bf 0204} (2002)
011, hep-th/0201224.
\bibitem{epz} N. Evans, M. Petrini, A. Zaffaroni.JHEP {\bf 0206} (2002) 004,
hep-th/0203203.
\bibitem{nsgub2}
V. Borokhov, S.S. Gubser, {\it Nonsupersymmetric Deformations of the Dual
of a Confining Gauge Theory},  hep-th/0206098.
\bibitem{bce1} J. Babington, D.E. Crooks, N. Evans, {\it A Non-supersymmetric
Deformation of the AdS/CFT Correspondence},  hep-th/0207076.
%
\bibitem{mn}J.M. Maldacena, C. Nunez, Phys. Rev. Lett. {\bf 86} (2001) 588,
hep-th/0008001.
%
\bibitem{ks} I.R. Klebanov, M.J. Strassler, JHEP {\bf 0008} (2000)
052, hep-th/0007191.
\bibitem{gb} C. Csaki, H. Ooguri, Y. Oz, J. Terning,  JHEP { \bf 9901}
(1999) 017,  hep-th/9806021.
%
\bibitem{karch} A. Karch, E. Katz, JHEP {\bf 0206} (2002) 043, hep-th/0205236.
%
\bibitem{hdcs} N. Evans, M. Petrini, JHEP {\bf 0111} (2001) 043,
hep-th/0108052.
%
\bibitem{shw} J.H. Schwarz, Nucl. Phys. {\bf B226} (1983) 269,
P.S. Howe, P.C. West,  Nucl. Phys. {\bf B238} (1984) 181.
%
\bibitem{neu}H.J. Kim, L.J. Romans, P. van Nieuwenhuize,
 Phys. Rev. {\bf D32} (1985) 389.
\bibitem{ma} M. Petrini, A. Zaffaroni, {\it The Holographic RG Flow to
Conformal and Non-Conformal Theory}, hep-th/0002172.



\end{thebibliography}
\end{document}